\documentclass[12pt]{article}
\pdfoutput=1
\usepackage[english]{babel}
\usepackage[latin1]{inputenc}
\usepackage{amsmath}
\usepackage{amssymb}
\usepackage{array}
\usepackage{url}
\usepackage{graphicx}
\usepackage{color}

\def\lsim{\mathrel{\raise.3ex\hbox{$<$\kern-.75em\lower1ex\hbox{$\sim$}}}}
\def\gsim{\mathrel{\raise.3ex\hbox{$>$\kern-.75em\lower1ex\hbox{$\sim$}}}}

\newcommand{\be}{\begin{equation}}
\newcommand{\ee}{\end{equation}}
\newcommand{\bea}{\begin{eqnarray}}
\newcommand{\eea}{\end{eqnarray}}
\newcommand{\ass}{\renewcommand{\arraystretch}{1}}
\newcommand{\asl}{\renewcommand{\arraystretch}{1.2}}

\renewcommand{\t}[1]{\theta_{#1}}

\newcommand{\dm}[1]{{\Delta m^2_{#1}}}

\begin{document}

\date{\mbox{ }}

\title{
\vspace{-3.1cm}
{\begin{flushright}
{\normalsize
ULB-TH/11-16\\
} 
\end{flushright}}
\vspace{1.2cm}
\bf Relations among neutrino observables in the light of a large $\theta_{13}$ angle\\[6mm]}

\author{Xiaoyong Chu, Mika\"el Dhen and Thomas Hambye
\\[3mm]
{\normalsize\it Service de Physique Th\'eorique,}\\[-0.15cm]
{\it\normalsize Universit\'e Libre de Bruxelles, 1050 Brussels, Belgium} \\[1.5mm]
}
\maketitle

\thispagestyle{empty}

\begin{abstract}
\noindent

The recent T2K and MINOS indications for a "large" $\theta_{13}$ neutrino mixing angle can be accommodated in principle by an infinite number of Yukawa flavour structures in the seesaw model.
Without considering any explicit flavour symmetry, there is an instructive exercise one can do:
to determine the simplest flavour structures which can account for the data with a minimum number of parameters, simply assuming these parameters to be uncorrelated.
This approach points towards a limited number of simple structures which show the minimum complexity a neutrino mass model must generally involve to account for the data.
These basic structures essentially lead to only 4 relations between the neutrino observables.
We emphasize that 2 of these relations, $|\sin \theta_{13}|=\frac{\tan\t{23}}{\cos\delta}\frac{1-\tan\t{12}}{1+\tan\t{12}}$ and $|\sin \theta_{13}| = \sin \theta_{12} R^{1/4}$, with $R\equiv \Delta m^2_{21}/\Delta m^2_{32}$, have several distinctive properties.
First, they hold not only with a minimum number of parameters, but also for complete classes of more general models. Second, any value of $\theta_{13}$ within the T2K and MINOS ranges can be obtained from these relations by taking into account small perturbations. Third, they turn out to be the pivot relations of models with approximate conservation of lepton number, which allow the seesaw interactions to induce observable flavour violating processes, such as $\mu \rightarrow e \gamma$ and $\tau \rightarrow \mu \gamma$.
Finally, in specific cases of this kind, these structures have the rather unique property to allow a full reconstruction of the seesaw Lagrangian from low energy data.

\end{abstract}

\newpage

\section{Introduction}

The T2K experiment has recently reported an indication for a "large" value of the third neutrino mixing angle, $0.02(0.03)< \sin^2 2\theta_{13} < 0.32(0.39)$ at $90\%$ CL, with central value $\sin^2 2\theta_{13}=0.12(0.14)$, for a normal (inverted) neutrino hierarchy. Soon after, the MINOS experiment \cite{Minos} reported somewhat smaller values 
$0< \sin^2 2\theta_{13} < 0.15(0.20)$ at $90\%$ CL, with a central value $\sin^2 2\theta_{13}=0.05$(0.07).
The impact of the knowledge of $\theta_{13}$ on the determination of the neutrino mass origin is not straightforward.
A value of $\theta_{13}$ within these ranges still allows many options. Moreover, in any specific neutrino mass generation scheme, the data could still be reproduced by a multidimensional continuum of flavour structures. For instance, in the seesaw model based on the existence of 3 heavy right-handed neutrinos, the knowledge of $\theta_{13}$ gives only one more constraint on a model which, on top of the 3 charged lepton masses, contain in full generality not less than 18 physical parameters (including 6 phases) in its basic Lagrangian
\begin{equation}
{\cal L} \owns -Y_{Nij} \bar{N}_i \tilde{\phi}^\dagger L_j -\frac{1}{2}M_{N_{ij}} \bar{N}_i N^c_j +h.c.\label{seesawL}
\end{equation}
with $\phi=(\phi^+,\,\phi^0)^T$ the standard model Higgs doublet, and $L_j=(\nu_{Lj}\, l^-_{Lj})$. 
This has to be compared with the number of neutrino observables. Four of them have been experimentally determined, $\Delta m^2_{31}$, $\Delta m^2_{12}$, $\theta_{12}$, $\theta_{23}$, see Ref.~\cite{schwetz,Lisi} for recent global fits of neutrino data. Together with $\theta_{13}$, and, possibly in a relatively near future, with the CKM type phase $\delta$ (especially if $\theta_{13}$ is large), and the $0\nu2\beta$ effective mass, $m_{ee}$, one could have access to 7 combinations of these parameters.
Even with 2 right-handed neutrinos, which is an allowed possibility, there is still a rather large number of physical parameters in the basic Lagrangian, 11 parameters (including 3 phases) on top of the charged lepton masses.
In this context no doubt that many flavour symmetries could in principle account for the data. 
Without considering any explicit one \cite{flavoursym}, an instructive exercise one can do is to determine
the structures which can account for the data, assuming the parameters of Eq.~(\ref{seesawL}) to be uncorrelated and a hierarchy among them, so that only a minimum number of them  \cite{Frampton:02a,Guo:02a} effectively matters. 
Barring in this way the infinity of possible special relations or cancellations among the parameters, which in principle could be considered, one systematically obtains basic structures which can account in a simple way for the data. These structures display
the minimum complexity a neutrino mass model must generally involve to accommodate the data.

\section{Basic relations between neutrino observables}

From the analysis of Ref.~\cite{BHR}, it turned out that a minimum of 4 "effective" real parameters is necessary, which in turn means that 2 relations between the 6 real neutrino mass matrix observables are predicted, relating $\theta_{13}$ and $m_{ee}$ with the other oscillation observables. 
The number of these predictions is very limited. Only 5 relations, between $\theta_{13}$ and the other oscillation parameters, turned out to be possible.
Three of them, called A (first considered in \cite{Frampton:02b,Raidal:02a}), B and E, predict a sizable value for $\theta_{13}$, which could be in agreement with a large value of $\theta_{13}$, along the new T2K or MINOS values, see Table 1.\footnote{As for cases C and D of Ref.~\cite{BHR}, they give $\sin^2 2\theta_{13}=0.0076 \pm 0.0007 $ and $\sin^2 2\theta_{13}= 0.000028\pm 0.00034$ at 1$\sigma$. Note that, in Ref.~\cite{BHR},  3 other structures were also found, called Fa, Fb and Fc. Fa gives a far too large value of $\theta_{13}$. Fc gives a large central value, with nevertheless sizable errors, $\sin^2 2\theta_{13}= (0.25\pm 0.03)/\cos^2\delta$,  which might be worth to look at in a closer way.
As for Fb it predicts a small value, $\sin^2 2\theta_{13}= (0.0007\pm 0.004)/\cos^2\delta$.} Another one, we call G, and which predicts $|\sin \theta_{13} |= \sin \theta_{12} R^{1/4}$ with $R\equiv \Delta m^2_{21}/\Delta m^2_{32}$, 
was put apart in Ref.~\cite{BHR}, on the basis of the Chooz upper bound on $\theta_{13}$. However, given the new estimation of the reactor neutrino flux of Ref.~\cite{newflux}, and the somewhat confusing situation it leads to, see e.g.~Ref.~\cite{schwetz}, one could wonder if it is still so firmly excluded. Taken at face value, it is perfectly within the range allowed by T2K, but excluded by other constraints at the $\sim 3 \sigma$ level \cite{schwetz,Lisi}.
In any case, given the large value of $\theta_{13}$ it predicts, it is fully testable in the near future.  
Moreover, as explained below, small perturbations of the structures leading to this prediction can easily lead to smaller values of $\theta_{13}$.
Note that all cases provide a way to solve the "large $\theta_{23}$-small $R$" problem without fine tuning, and the case "E" provides also a natural way to obtain for $\theta_{12}$ a large deviation from $\pi/4$
with inverted hierarchy.
\asl
\begin{table}[t]
  \centering
  \[\begin{array}{|c||c|c|c|c|c|c|} 
    \hline
    &
   | \sin\t{13}| & \sin^2 2\theta_{13}\,\,(1\sigma) &
    |m_{ee}|/m_{atm} \,\,\text{(eV)}&
    \text{SSC}  &
    U_l \\
    \hline\hline
    \text{A1}&
    \displaystyle\rule[-0.4cm]{0cm}{1.1cm}
    \frac{1}{2}\tan\t{23}\sin2\t{12}\,\sqrt{R} & 0.027\pm 0.007 &
    \sin^2\t{12}\sqrt{R}&
    \text{2} &
    \mathbf{1} \\
    \hline
    \text{B1} &
    \displaystyle\rule[-0.4cm]{0cm}{1.1cm}
    \frac{1}{2}\tan\t{23}\tan2\t{12}\,(R\cos2\t{12})^{1/2} &0.072\pm0.019 &
    0 &
    \text{3} &
    \mathbf{1} \\
    \hline
    \text{E1} &
    \displaystyle\rule[-0.4cm]{0cm}{1.1cm}
    \frac{\tan\t{23}}{\cos\delta}\frac{1-\tan\t{12}}{1+\tan\t{12}} &  \frac{0.158\pm 0.046}{\cos^2 \delta}
    &
    2\cot\t{23}\sin\t{13} &
    \text{2, 3} &
    \mathbf{1},R_{12},R_{23} \\
    \hline
    \text{G} &
    \displaystyle\rule[-0.4cm]{0cm}{1.1cm}
    \sin \theta_{12} R^{1/4} & 0.194 \pm 0.010
    &
   0&
    \text{2} &
    1\\
    \hline

  \end{array}\]
  \caption{Summary of the possible correlations between $\t{13}$,
  $m_{ee}$ and $\t{23}$, $\t{12}$, $\delta$, 
  $R$, to leading order of the expansion
  in $R$ (with $m_{\text{atm}}\equiv\sqrt{|\dm{32}|}$). The 
  column ``SSC'' gives 
  the number of right-handed neutrinos involved in the see-saw
  realization of each case. The "$U_l$" column refers to the charged lepton rotation performed, see Ref.~\cite{BHR}. All cases involve a normal hierarchy except case E. Cases A2, B2, E2, are obtained from A1, B1, E1, with 
  the replacements $\tan\t{23} \rightarrow \cot\t{23}$ and 
  $\cos \delta \rightarrow - \cos\delta$.
  The non-vanishing predictions for $m_{ee}$ give $m_{ee}=(0.0027\pm0.0002)\,\text{eV}$   and $m_{ee}=\frac{0.019 \pm 0.002}{\cos\delta}\, \text{eV}$ for the A and E cases respectively. Only the inverted hierarchy pattern "E" gives a large $m_{ee}$ value, as expected \cite{Feruglio:02a}. All numerical results have been obtained using the $\theta_{12}$, $\theta_{23}$, $\Delta m^2_{13}$, $\Delta m^2_{12}$ global fit values of Ref.~\cite{schwetz}. Using instead the global fit results of Ref.~\cite{Lisi} lead to essentially the same result apart for a substantial difference for $\theta_{23}$ which reduces the value of $\sin^2 2 \theta_{13}$ for case A, B, and E by $\sim20 \%$.} 
  \label{tab:results}
\end{table}
\ass

In this letter, beside updating in Table 1 the predictions of cases A, B and E, we consider more extensively the inverted hierarchy E scenario and normal hierarchy G scenario, exact or perturbed.
To this end, to count the number of effective parameters, it is convenient to define the Dirac mass matrix, $m_{D{ij}}=Y_{Nij} v$ as 
\begin{equation}
m_D=y\,A\,,
\end{equation}
where $y_{ij}=y_i \delta_{ij}$ and $(AA^\dagger)_{ii}=1$ with $i=1,2(1,2,3)$ for 2 (3) right-handed neutrinos.
In terms of the $A$ matrix, the light neutrino mass matrix can be written as $M_\nu=-A^T \mu^{-1} A$ with $\mu^{-1}\equiv y M_N^{-1} y$ (A is adimensional while $\mu$ has the dimension of an inverse mass). The number of effective parameters is the number of parameters contained in $\mu$ and A.
The 4 effective parameter structures which lead to case A, B, with 2 or 3 right-handed neutrinos, can be found in Ref.~\cite{BHR}. With such limited number of parameters they involve at most one CP violating phase.

As for the E scenario there are many structures which can lead to it with 2 or 3 right-handed neutrinos, as shown in Ref.~\cite{BHR}.
Structures not made explicit there have
\begin{equation}
 A = 
\begin{pmatrix}
c & s & 0 \\
\alpha e^{i\phi} & \beta &\gamma 
\end{pmatrix} \,  ,
\quad\text{with}\quad
 \mu =\mu_0
\begin{pmatrix}
0 & 1 \\
1 & 0
\end{pmatrix}\,,
\label{E1structure}
\end{equation}
leading to E1 prediction, or
\begin{equation}
  A = 
\begin{pmatrix}
c & 0 & s \\
\alpha e^{i\phi} & \beta &\gamma 
\end{pmatrix}  ,
\quad\text{with}\quad
 \mu =\mu_0
\begin{pmatrix}
0 & 1 \\
1 & 0
\end{pmatrix}\,,
\label{E2structure}
\end{equation}
leading to E2 prediction.
Both structures can be obtained from each other via the exchange $\nu_\mu \leftrightarrow \nu_\tau$ and $\theta_{23} \leftrightarrow \pi/2-\theta_{23}$.\footnote{Remark that that the E relation can also be obtained \cite{BHR} if we assume hierarchies directly among the entries of the neutrino mass matrix, what makes sense in the framework of the type-II seesaw model, since in this model this matrix is directly proportional to the Yukawa coupling matrix of the scalar triplet.}

The structures which lead to prediction G,  considered in Ref.~\cite{BHR}, are obtained with 2 (dominating) right-handed neutrinos and 
\begin{equation}
  \label{AH}
  A = 
\begin{pmatrix}
0 & s & c \\
c' & s' e^{i\phi} & 0
\end{pmatrix} 
\quad\text{or}\quad
  A = 
\begin{pmatrix}
0 & s & c \\
c' & 0 & s' e^{i\phi}
\end{pmatrix}\, ,
\quad\text{with}\quad
 \mu = \mu_0
\begin{pmatrix}
0 & \varepsilon \\
\varepsilon & 1
\end{pmatrix} \,.
\end{equation}
It splits into 2 structures according to both A matrices in Eq.~(\ref{AH}). Here too both possibilities can be obtained from each other via the exchange $\nu_\mu \leftrightarrow \nu_\tau$ and $\theta_{23} \leftrightarrow \pi/2-\theta_{23}$.
On top of these structures, it turns out that there is another one leading to prediction G with 4 effective parameters:
\begin{equation}
  \label{AH2}
  A = 
\begin{pmatrix}
0 & s & c \\
\alpha & \beta &\gamma e^{i\phi}
\end{pmatrix} 
\quad\text{with}\quad
 \mu =\mu_0
\begin{pmatrix}
0 & 1 \\
1 & 0
\end{pmatrix}\,.
\end{equation}
While the structures of Eq.~(\ref{AH})  involve a hierarchical pattern of right-handed neutrino masses, the one of Eq.~(\ref{AH2}) displays a degenerate spectrum. This doesn't prevent these 2 cases from leading to the same G prediction for $\theta_{13}$ and $m_{ee}$.

\section{The inverted hierarchy case}

\subsection{The E relation as a generic 2 right-handed neutrino pre\-diction}

The E relation, unlike the G one, is allowed experimentally at the level of about $1\sigma$ (or less using the global fit data of Ref.~\cite{Lisi}).
As shown in Ref.~\cite{BHR}, it can be obtained from many different patterns with 2 or 3 right-handed neutrinos. Here we would like to emphasize the fact that it is generic of a complete class of seesaw models with 2 right-handed neutrinos.
Consider any general 2 right-handed neutrino structure,\footnote{For a study of the 2 right-handed neutrino case, see Ref.~\cite{Ibarra:2005qi}. Note also that it is known that a 2 by 2 $\mu$ block with one dominant entry and non-zero determinant can account for the data on atmospheric and solar oscillation parameters  \cite{Smirnov:93a,King:98a}.}
i.e.~a heavy and light neutrino mass matrix
in the $(\nu_L N_1 N_2)$ basis of the form
\begin{equation}
{\cal M}=
\begin{pmatrix}
0 & Y_N^T v & Y'^T_N v\\
Y_N v & M_{N_{11}} & {M_{N_{12}}}\\
Y'_N v & M_{N_{12}}   & M_{N_{22}}
\end{pmatrix}\,.
\label{mnutotal}
\end{equation}
with $Y_N$ ($Y'_N$) a $1\times 3$ matrix denoting the Yukawa couplings of the first (second) right-handed neutrino.
Eq.~(\ref{mnutotal}) gives the light neutrino mass matrix
\begin{equation}
M_\nu=- v^2 [Y_N^T (M_N^{-1})_{11} Y_N +(Y^T_N (M_N^{-1})_{12} Y'_N+Y'^T_N (M_N^{-1})_{12} Y_N) + Y'^T_N (M_N^{-1})_{22} Y_N']\,.
\label{Mnudecomp}
\end{equation}
It turns out that if one of the right-handed neutrino, say $N_1$, feebly couples to the $\tau$ (or $\mu$) flavour, $Y_{N\tau}\simeq0$ ($Y_{N\mu}\simeq0$), and if in this general neutrino mass matrix formula, the $(M_N^{-1})_{22}$ term has a negligible contribution, one readily obtains the E1 (E2) prediction.
This stems from the fact that, as observed in Ref.~\cite{GHHH}, if  the $(M_N^{-1})_{22}$ term in Eq.~(\ref{Mnudecomp}) is negligible, one can rewrite $M_\nu$ as
\begin{equation}
M_\nu=- v^2 (Y_N^T (M_N^{-1})_{12} Y''_N+Y''^T_N (M_N^{-1})_{12} Y_N) \,,
\label{mnuspecial}
\end{equation}
with $Y''_N=Y'_N+\frac{(M_N^{-1})_{11}}{2(M_N^{-1})_{12}} \, Y_N$.
A diagonalisation of the neutrino mass matrix with such a structure, for a normal neutrino hierarchy, gives for the $\tau$ Yukawa coupling \cite{GHHH} 
\begin{equation}
Y_{N\tau}=\frac{y}{\sqrt{2}}\,(\sqrt{1+\rho} \, U_{32}^* + \sqrt{1-\rho} \,U_{31}^*)\,,
\label{YNtau}
\end{equation}
where $U_{31}=s_{12} s_{23} e^{-i\alpha}- c_{12} c_{23} s_{13} e^{i(\delta+\alpha)}$ and $U_{32}=-c_{12} s_{23}e^{i\alpha}- s_{12} c_{23} s_{13} e^{i(\delta+\alpha)}$ are the usual PMNS matrix elements, with $\delta$ the CKM type phase and $\alpha$ the unique Majorana phase one has with 2 right-handed neutrinos. In Eq.~(\ref{YNtau}),  $\rho=(\sqrt{1+R}-1)/(\sqrt{1+R}+1)$ and $y$ is the $Y_N$ normalization factor, i.e.~$y=\sqrt{\sum_i |Y_{Ni}|^2}$.
Put the other way one has consequently
\begin{equation}
\sin \theta_{13}=- \frac{s_{23}}{c_{23}}\frac{
\sqrt{1+\rho}\,c_{12}e^{-i\alpha}\,\,-\,\sqrt{1-\rho}\,s_{12}\,e^{i \alpha}\,\,\,\,\,\,\,}{\sqrt{1-\rho}\,c_{12}e^{i(\alpha-\delta)}+\sqrt{1+\rho}\,s_{12}e^{-i(\alpha+\delta)}}+\sqrt{2}\,\frac{Y_{N\tau}}{y}\frac{1}{D}
\label{sinthetaYNtau}
\end{equation}
with $D=c_{23}({\sqrt{1-\rho}c_{12}e^{i(\alpha-\delta)}+\sqrt{1+\rho}s_{12}e^{-i(\alpha+\delta)}})$.
In other words, if $|Y_{N\tau}| = 0$, 
\begin{equation}
|\sin \theta_{13}|=\frac{\tan \theta_{23}}{\cos \delta} \Big(\frac{\sqrt{1+\rho}-\sqrt{1-\rho}\,\tan \theta_{12}}{\sqrt{1-\rho}+\sqrt{1+\rho}\,\tan \theta_{12}}\Big)
\label{finaltheta13tau}
\end{equation}
Eq.~(\ref{finaltheta13tau}) gives the $\sin^22 \theta_{13}= (0.158\pm 0.046)/\cos^2 \delta$ value quoted in Table 1 and, at first order in $R$,  gives nothing but the E1 relation 
$|\sin \theta_{13}|=\frac{\tan\t{23}}{\cos\delta}\frac{1-\tan\t{12}}{1+\tan\t{12}}$ which gives basically the same value $\sin^22 \theta_{13}= (0.152\pm 0.045)/\cos^2 \delta$.
Note that the factor $1/\cos \delta$ in Eq.~(\ref{finaltheta13tau}) follows from the non-linear relation we get in this case between the Majorana and CKM type phase, and is a good approximation of the exact result\footnote{Actually we get $\sin \theta_{13}=-\tan\theta_{23} \cos 2 \theta_{12}/(\sqrt{\cos^2\delta+\cos^2 2 \theta_{12} \sin^2 \delta}+\cos\delta\sin2\theta_{12})$.} for $\cos \delta \sim 1$. To have a rather small CP-phase is necessary to avoid a too large $\theta_{13}$ value. For example a value of $\sin^2 2\theta_{13}<0.20$ requires
$\cos \delta \gtrsim 0.85$, i.e.~$|\delta| \lesssim 30^\circ$.

If instead of assuming $Y_{N\tau}\simeq 0$ one assumes $Y_{N\mu}\simeq 0$ one obtains in the same way the E2 relation instead of the E1 one, differing just by
  the replacements $\tan\t{23} \rightarrow \cot\t{23}$ and 
  $\cos \delta \rightarrow - \cos\delta$.

In other words any 2 right-handed neutrino structure which considers one right-handed neutrino, say $N_1$, with a negligible coupling to the tau or mu flavour, and has a subdominant $Y'^T_N(M_N^{-1})_{22} Y'_N$ term in Eq.~(\ref{Mnudecomp}) leads to the E1 and E2 relation respectively. The latter condition is satisfied generically as soon as $Y'_N$ is small and/or $M_{N_{11}}$ is small (or more generally with a hierarchy among the $M_N$ entries, with in all cases a non-vanishing off-diagonal entry $M_{N_{12}}$). In turn,  a prediction of this general class of models is a sizable value of the $0\nu2\beta$ effective mass, as given in Table 1. 
The structures of Eqs.~(\ref{E1structure})-(\ref{E2structure}) are 4 effective parameter subcases of this general class of models, i.e. they lead in particular to a vanishing $Y'^T(M_N^{-1})_{22} Y'$  term. 

Let us emphasize that if $\sin^2 2\theta_{13}$ is precisely determined in the future with a smaller value than predicted by Eq.~(\ref{finaltheta13tau}), what is not clear at the present stage, still a small perturbation of this case can account for the data. For example, for $\alpha=\delta=0$, the value $\sin^2 2 \theta_{13}=0.08$ is obtained in Eq.~(\ref{sinthetaYNtau}) with  $|\frac{Y_{N\tau}}{y}|\simeq 0.05$, which is indeed a small perturbation, i.e.~$Y_{N\tau}<< Y_{Ne,\mu}$. The E relation  can therefore be considered more generally as a pivot relation around which relevant values of $\theta_{13}$ are obtained naturally.
In other words the small value of $\theta_{13}$ in these models can be traced back from the simple fact that one of the right-handed neutrino couples less to the $\tau$ or $\mu$ flavour than to the other flavours.
Of course values of $\sin^2 2 \theta_{13}$ as low as for example 0.02 would require a cancelation in Eq.~(\ref{sinthetaYNe}) at the one per ten level, which is not unbearable, but somehow goes against the original point of view of Ref.~\cite{BHR}.

\subsection{Approximate conservation of lepton number and flavour violating rates}

In Ref.~\cite{GHHH}, has been considered a general class of 2 right-handed neutrino models 
based on the approximate conservation of lepton number symmetry.\footnote{Related structures have been considered in Refs.~\cite{Wyler:1982dd,Mohapatra:1986bd,Branco:1988ex,Raidal:2004vt,Kersten:2007vk,Abada:2007ux,ibarra}.} 
They are based in Eq.~(\ref{mnutotal}) on the hierarchy $Y'_N<<Y_N$ and $M_{N_{12}}>> M_{N_{11}},M_{N_{22}}$. That can be justified by assigning  $L(N_1)=1$, $L(N_2)=-1$ and, to induce the neutrino masses, adding small $L$-violating perturbations $Y'_N$, $M_{N_{ii}}$.
This possibility has the interesting phenomenological virtue to decouple the size of the dimension 6 coefficients from the size of the dimension 5 (neutrino mass) ones.
The coefficients of the low energy dimension 6 operator induced by the seesaw interactions, ${\cal L}_{eff} \owns c^{d=6}_{\alpha \beta} \bar{L}_\alpha \tilde{\phi} \,i \partial \hspace{-2.2mm} \slash(\tilde{\phi}^\dagger L_\beta)$,
 govern the size of the lepton flavour violating and L-conserving processes induced, with $c^{d=6}_{\alpha\beta}=(Y_N^\dagger (M_{N_{12}})^{-2} Y_N)_{\alpha\beta}+{\cal O}(Y'_N)$.
As a result, although the neutrino masses are tiny, observable flavour violating rates can be obtained. This requires that some of the $Y_N$ couplings are large enough and that the right-handed neutrino mass scale $M_{N_{12}}$ is not too far above the TeV scale, see Ref.~\cite{GHHH}. Moreover this possibility is of the minimal flavour violation type (MFV) in the sense that the knowledge of the dimension 5 neutrino mass matrix flavour structure allows to determine completely the flavour structure of the dimension 6 coefficients. 
In this class of models, since the term quadratic in $Y'_N$ is of second order in the L-violating perturbations, the last term of Eq.~(\ref{Mnudecomp}) can be neglected and  the structure of Eq.~(\ref{mnuspecial}) holds \cite{GHHH} . Therefore,  all models considered in Ref.~\cite{GHHH} for an inverted hierarchy lead to Eq.~(\ref{sinthetaYNtau}) (or similar one for the $\mu$ flavour). In other words the relation of Eq.~(\ref{finaltheta13tau}) is the pivot relation of the models which, based on approximate conservation of lepton number, allow 
observable flavour violating rates, with inverted hierarchy.\footnote{Extra possibilities to induce the baryon asymmetry of the Universe through leptogenesis in approximately L conserving setups have been discussed in Refs.~\cite{Blanchet:2009kk,Deppisch:2010fr}.}

\subsection{Full reconstruction of the seesaw Lagrangian from low energy data}

The 4 effective parameter cases of Eqs.~(\ref{E1structure})-(\ref{E2structure}) are special cases of the MFV models considered in Ref.~\cite{GHHH}. For the situation which can give large flavour violating rates, i.e. relatively low $M_N$ and large $y$, the general form of the dimension 6 coefficient in the $e,\,\mu,\,\tau$ basis is simply 
\begin{equation}
c^{d=6}=\frac{y^2}{M^2_{N_{12}}}\,\begin{pmatrix}
c^2 & c s& 0 \\
c s & s^2& 0\\
0 & 0  & 0
\end{pmatrix}\,
\quad\text{and}\quad
c^{d=6}=\frac{y^2}{M^2_{N_{12}}}\,\begin{pmatrix}
c^2 & 0 & c s \\
0 & 0  & 0\\
c s & 0 & s^2
\end{pmatrix}\,
\label{dim_6}
\end{equation}
for Eqs.~(\ref{E1structure}) and~(\ref{E2structure}) respectively.
As a result, in this case Eq.~(\ref{E1structure}) leads to 
suppressed $\tau\rightarrow e$ and $\tau \rightarrow \mu$ transition rates but the $\mu\rightarrow e$ one can be sufficiently large to saturate the present experimental bound on $\Gamma(\mu \rightarrow e \gamma)$.
Similarly Eq.~(\ref{E2structure}) leads to 
suppressed $\tau\rightarrow \mu$ and $\mu \rightarrow e$ transition rates but the $\tau \rightarrow e$ one can be sufficiently large to saturate the present experimental bound on $\Gamma(\tau \rightarrow e \gamma)$.
Moreover it is interesting to point out that in these cases, neglecting higher order contributions in $(Y_N v/M_N)^2$, the ratio of the $l \rightarrow l' \gamma$ and $l \rightarrow l' \, l''^+ l''^-$ induced rates (with $l''=e,\,\mu,\,\tau$) depends only on the value of the degenerate right-handed neutrino mass. This results from the fact that both rates are proportional to the same $l-l'$ dimension 6 coefficient. 
Using the analytical results for each of these branching ratios \cite{Pilaftsis}, Fig.~1 shows the value of the ratios obtained. They decrease if $M_N$ increases (for $M_N\gtrsim 200$~GeV).
A measure of this ratio would allow consequently to determine the right-handed neutrino mass scale $M_{N_{12}}$. Combined with the fact that the full flavour structure of the MFV models (with $M_{Nii}=0$ as in Eq.~(\ref{AH2})) can be reconstructed from the knowledge of the neutrino mass matrix \cite{GHHH}, this means that the scenarios of Eqs.~(\ref{E1structure}) and (\ref{E2structure}) have the property that their Lagrangian is fully reconstructible from the low energy data. For instance if the flavour structure and the right-handed neutrino mass scale are known, all parameters are known except the normalization factors of the Yukawa couplings. But these ones can be determined too. The $Y_N$ Yukawa coupling normalization factor, $y$, can be straightforwardly determined from the experimental value of the branching ratio of a single flavour violating process, Eq.~(\ref{dim_6}). Hence, the $Y'_N$ Yukawa coupling normalization factor, $y'\equiv \sqrt{\sum |Y'_{Ni}|^2}$, can be determined from the light neutrino mass scale, $M_\nu=-v^2(Y_N^T (M_N^{-1})_{12}Y'_N +Y'^T_N (M_N^{-1})_{12}Y_N)$.
\begin{figure}
\begin{tabular}{c}
\includegraphics[width=5.cm]{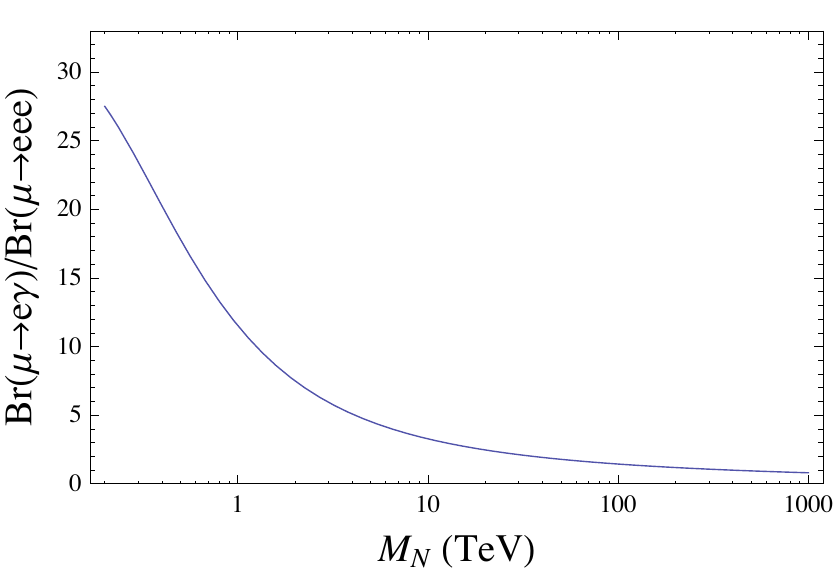}
\includegraphics[width=5.cm]{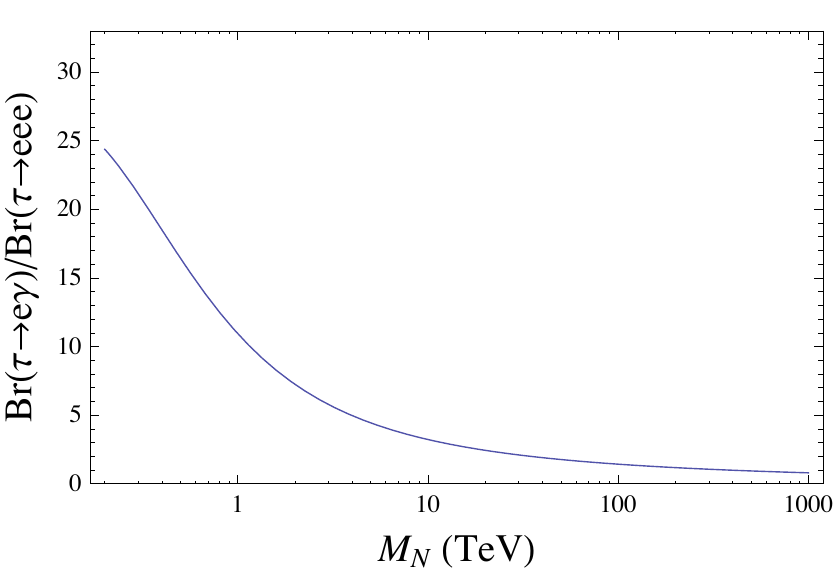}
\includegraphics[width=5.cm]{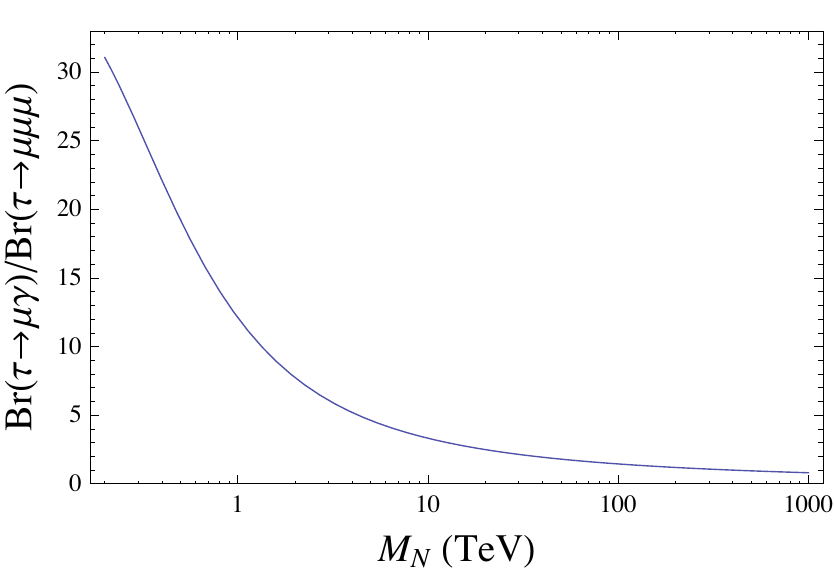}
\end{tabular}
\caption{Values of the $Br(l\rightarrow l' \gamma)/ Br(l \rightarrow l'l'l')$ ratios as a function of the right-handed neutrino mass scale for seesaw models with degenerate right-handed neutrinos.} \label{efficiencies}
\end{figure}

The same conclusions hold for the model where an extra small $Y_{N\tau}$ or $Y_{N\mu}$ perturbation is added to account for a possibly smaller value of $\theta_{13}$, Eq.~(\ref{sinthetaYNtau}), in case one also gets sizable rates for the other two $l \rightarrow l'$ channels, since the dimension-6 operator receives now contributions from these perturbations. These rates are suppressed by a $(Y_{N\mu,\tau}/y)^2$ but are still in principle observable.
As for $m_{ee}$ and the $Br(\tau\rightarrow l \gamma)/Br(\mu\rightarrow e \gamma)$ ratios, they can be found in Ref.~\cite{GHHH} as a function of $\theta_{13}$, the CKM type phase and the Majorana phase. 
These models are unique in involving so few parameters to reproduce the data
and allowing at the same time for observable rare lepton flavour violating processes, and for a reconstruction of the total seesaw Lagrangian. In turn, additional possibilities of producing the right-handed neutrinos at the LHC, from the large $Y_N$ couplings, have been studied in related models in Ref.~\cite{ibarra}. The full reconstruction procedure above predicts the scale where, along this scenario, the right-handed neutrinos have to be found at accelerators.

It must be stressed that the predictions for the $Br(l\rightarrow l' \gamma)/ Br(l \rightarrow l'l'l')$ ratios
in Fig.~1 hold for any model with quasi-degenerate right-handed neutrinos, simply because, if there is only one right-handed neutrino mass scale, the Yukawa couplings cancel in the ratio, no matter what they are (neglecting higher order contributions in $(Y_N v/M_N)^2$).
A determination of the right-handed neutrino mass is therefore feasible in this way for any model of this kind.
Alternatively the ratio between the $\mu$ to $e$ conversion rate in atomic nuclei and the $\mu\rightarrow e \gamma$ branching ratio could be used 
to determine the degenerate right-handed neutrino mass scale in the same way.

\section{The normal hierarchy case}

The G relation, which holds for a normal hierarchy, has properties very similar to the inverted hierarchy E relation.
In the 2 right-handed neutrino case it can be obtained in the same way as the E relation.
For instance if  the $(M_N^{-1})_{22}$ term in Eq.~(\ref{Mnudecomp}) is negligible, so that one can rewrite $M_\nu$ as in Eq.~~(\ref{mnuspecial}), in the normal hierarchy case we get the relation 
\cite{GHHH},\footnote{See also Ref.~\cite{Raidal:2004vt} which considers the normal hierarchy case with $M_{N_{11}}=M_{N_{22}}=0$.} 
\begin{equation}
Y_{Ne}=\frac{y}{\sqrt{2}}\,(\sqrt{1+\rho} \, U_{13}^* + \sqrt{1-\rho} \,U_{12}^*)\,,
\label{YNe}
\end{equation}
with $U_{13}=\sin \theta_{13} e^{-i\delta}$ and $U_{12}=\sin \theta_{12} \cos \theta_{13} e^{i\alpha}$,  the corresponding PMNS matrix elements. Here too $\delta$ is the CKM type phase and $\alpha$ is the unique Majorana phase one has with 2 right-handed neutrinos. In Eq.~(\ref{YNe}), $\rho=(\sqrt{1+R}-\sqrt{R})/(\sqrt{1+R}+\sqrt{R})$ and, as above, $y$ is the $Y_N$ normalization factor, i.e.~$y=\sqrt{\sum_i |Y_{Ni}|^2}$.
Put the other way,
\begin{equation}
\sin \theta_{13}=-\frac{\sqrt{1-\rho}}{\sqrt{1+\rho}}\sin \theta_{12} \cos \theta_{13} e^{-i(\alpha+\delta)} +\frac{Y_{Ne}}{y}\frac{\sqrt{2}}{\sqrt{1+\rho}}e^{-i\delta}\,,
\label{sinthetaYNe}
\end{equation}
or in other words, if $|Y_{Ne}| \simeq 0$, 
\begin{equation}
|\sin \theta_{13}|=\frac{\sqrt{1-\rho}}{\sqrt{1+\rho}}\sin \theta_{12} \cos \theta_{13} \,.
\label{finaltheta13}
\end{equation}
Eq.~(\ref{finaltheta13}) gives the $\sin^22 \theta_{13}=0.194\pm 0.010$ value quoted in Table 1 and, at first order in $R^{1/4}$ and $\sin \theta_{13}$,  gives nothing but the relation $|\sin \theta_{13}|=\sin \theta_{12}  R^{1/4}$.

Let us emphasize here too that if $\sin^2 2\theta_{13}$ is precisely determined in the future with a smaller value than predicted by Eq.~(\ref{finaltheta13}), as favored experimentally, still a small perturbation of this case can account for the data. For example $\sin^2 2 \theta_{13}=0.08$ is obtained in Eq.~(\ref{sinthetaYNe}) with $|\frac{Y_{Ne}}{y}\frac{\sqrt{2}}{\sqrt{1+\rho}}| \simeq 0.11$, that is to say $|\frac{Y_{Ne}}{y}|\simeq 0.10$, which is indeed a small perturbation, i.e.~$Y_{Ne}<< Y_{N\mu,\tau}$. The relation $|\sin \theta_{13}|=\sin \theta_{12}  R^{1/4}$ can therefore be considered as the first order relation around which relevant values of $\theta_{13}$ are obtained naturally.
In other words the small value of $\theta_{13}$ in these models can be traced back from the fact that one of the right-handed neutrino couples less to the $e$ flavour than to the $\mu$, $\tau$ ones.
This is just the opposite of the inverted hierarchy case which requires a suppressed coupling in the $\mu$ or $\tau$ channel.

The G relation is also the pivot relation of the models with approximate conservation of lepton number and normal hierarchy, since here too the term quadratic in $Y'_N$ is of second order in the L-violating perturbations, so that the last term of Eq.~(\ref{Mnudecomp}) can be neglected and  the structure of Eq.~(\ref{mnuspecial}) holds.
As a result this is also the structure which allows large lepton flavour violating rates, with a dimension 6 coefficient different from the ones with an inverted hierarchy,
\begin{equation}
c^{d=6}=\frac{y^2}{M^2_{N_{12}}}\,\begin{pmatrix}
0 & 0& 0 \\
0 & s^2& sc\\
0 & s c  & c^2
\end{pmatrix}\,,
\end{equation}
for $Y_{Ne}=0$ and subleading terms in the first row and column for $Y_{Ne}\neq0$.
As a result the $\mu\rightarrow e$ and $\tau \rightarrow e$ transition rates are subleading, because suppressed by a $(Y_{Ne}/y)^2$ factor but still in principle observable. The $\tau\rightarrow \mu$ transition can be sufficiently large to saturate the present experimental bound on $\Gamma(\tau \rightarrow \mu \gamma)$ or $\Gamma (\tau \rightarrow \mu\, l^+ l^-)$ with $l=e,\,\mu,\,\tau$.

A full reconstruction of the seesaw Lagrangian for the structure of Eq.~(\ref{AH2}), with or without an additional $Y_{N\tau}$ or $Y_{N\mu}$ entry, is possible in the same way as above for the inverted hierarchy, extracting the right-handed neutrino mass scale from Fig.~1.c above, determining $y$ from the size of a single flavour violating process and the normalization of the $N_2$ Yukawa coupling from the neutrino mass scale, and, in the way explained in Ref.~\cite{GHHH}, extracting the full flavour structure of the model from the knowledge of the 7 neutrino mass matrix observables.

\section{Summary}

In the light of the new indications for a "large" $\theta_{13}$ angle \cite{T2K,Minos}, a limited number of basic structures emerges, accounting for the data from a minimal number of uncorrelated seesaw parameters.
The relations among neutrino observables they lead to are given in Table 1. The structures A, B and E appear to be the most favored by the data. 
The inverted hierarchy structure E is generic of many different structures with 2 or 3 right-handed neutrinos.
The G structure is also interesting, as it predicts a value of $\theta_{13}$ around the experimentally allowed ranges, and can take any value within these ranges, if perturbed. The predictions the E and G structures lead to, Eqs.~(\ref{finaltheta13tau}) and (\ref{finaltheta13}), or more generally Eq.~(\ref{sinthetaYNtau}) and (\ref{sinthetaYNe}) when perturbed, are in particular generic of a complete class of models dominated by 2 right-handed neutrinos.
They hold, exactly or in first approximation, for any viable seesaw structure based on 2 right-handed neutrinos, if one of the diagonal entries of the right-handed neutrino mass matrix has a negligible contribution to the light neutrino masses.
Interestingly, the E and G structures turn out to be
also the pivot relations of the 2 right-handed neutrino models which, with approximate conservation of lepton number, can reproduce the data. Therefore, they allow for observable 
lepton flavour violating transition rates, induced exclusively by the seesaw interactions.
Moreover they can be obtained for models which have the rather unique property to allow for a full reconstruction of the seesaw Lagrangian from low energy data. Of course such a specific subscenario with observable flavour changing processes is only one among many possible ones, but it shows that this full reconstruction is in principle feasible in specific cases.
Similarly the measurement of $\theta_{13}$, and possibly of $m_{ee}$, in precise agreement with one of the predictions A, B, E or G, wouldn't prevent the fact that still an infinity of seesaw structures (with more right-handed neutrinos, more parameters or other neutrino mass generation mechanisms) could lead to the same values, "by chance", but that would be suggestive enough to undertake further investigations.

\section*{Acknowledgements}
We thank B.~Gavela and T.~Schwetz for useful discussions.
This work is supported by 
the FNRS-FRS, the IISN and the Belgian Science Policy (IAP VI-11).
The results of this letter concerning the G structure can be found to a large extent in the Master thesis of MD, ULB, May 2011.

\bibliographystyle{JHEP}
\bibliography{bibliography.bib}

\end{document}